\documentclass{article}

\usepackage{arxiv}
\pdfoutput=1

\usepackage[utf8]{inputenc} 
\usepackage[T1]{fontenc}    
\usepackage{booktabs}       
\usepackage{amsfonts}       
\usepackage{nicefrac}       
\usepackage{microtype}      
\usepackage{graphicx}
\usepackage{doi}
\usepackage{enumitem}

\title{Visualization and Analysis of Wearable Health Data From COVID-19 Patients}

\author{ \href{https://orcid.org/0000-0002-7998-7930}{\includegraphics[scale=0.06]{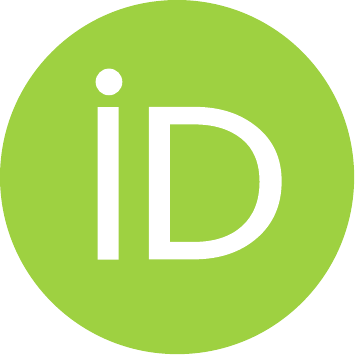}\hspace{1mm}Susanne K.~Suter}\thanks{Zurich University of Applied Sciences (ZHAW), Computational Health Lab, Waedenswil, Switzerland}
	\And
	\href{https://orcid.org/0000-0001-9640-8155}{\includegraphics[scale=0.06]{orcid.pdf}\hspace{1mm}Georg R.~Spinner}\footnotemark[1]
	\And
	\href{https://orcid.org/0000-0000-0000-0000}{\includegraphics[scale=0.06]{orcid.pdf}\hspace{1mm}Bianca Hoelz}\thanks{University Hospital Basel (USB), D\&ICT and CMO Office, Basel, Switzerland}
	\And
	\href{https://orcid.org/0000-0000-0000-0000}{\includegraphics[scale=0.06]{orcid.pdf}\hspace{1mm}Sofia Rey}\footnotemark[1]
	\And
	\href{https://orcid.org/0000-0000-0000-0000}{\includegraphics[scale=0.06]{orcid.pdf}\hspace{1mm}Sujeanthraa  Thanabalasingam}\thanks{University Hospital Basel (USB), Internal Medicine, Basel, Switzerland}
	\And
	\href{https://orcid.org/0000-0002-1693-9851}{\includegraphics[scale=0.06]{orcid.pdf}\hspace{1mm}Jens Eckstein}\thanks{University Hospital Basel (USB), Internal Medicine and D\&ICT, Basel, Switzerland}
	\And
	\href{https://orcid.org/0000-0002-5678-4110}{\includegraphics[scale=0.06]{orcid.pdf}\hspace{1mm}Sven Hirsch}\footnotemark[1]
}

\renewcommand{\shorttitle}{Wearable Data From COVID-19 Patients}

\hypersetup{
pdftitle={Wearable Health Data From COVID-19 Patients},
pdfsubject={Wearable-Health-Data-Vis},
pdfauthor={Susanne K.~Suter, Georg R.~Spinner, Bianca Hoelz, Sofia Rey, S. Thanabalasingam, Jens Eckstein, Sven Hirsch},
pdfkeywords={Wearable vital signs, COVID-19 patients, visualizations},
}

\begin{document}
\maketitle

\begin{abstract}
Effective visualizations were evaluated to reveal relevant health patterns from multi-sensor real-time wearable devices that recorded vital signs from patients admitted to hospital with COVID-19. Furthermore, specific challenges associated with wearable health data visualizations, such as fluctuating data quality resulting from compliance problems, time needed to charge the device and  technical problems are described. As a primary use case, we examined the detection and communication of relevant health patterns visible in the vital signs acquired by the technology. Customized heat maps and bar charts were used to specifically highlight medically relevant patterns in vital signs. A survey of two medical doctors, one clinical project manager and seven health data science researchers was conducted to evaluate the visualization methods. From a dataset of 84 hospitalized COVID-19 patients, we extracted one typical COVID-19 patient history and based on the visualizations showcased the health history of two noteworthy patients. The visualizations were shown to be effective, simple and intuitive in deducing the health status of patients. For clinical staff who are time-constrained and responsible for numerous patients, such visualization methods can be an effective tool to enable continuous acquisition and monitoring of patients’ health statuses – even remotely. 
\end{abstract}


\section{Introduction}

\subsection{Patient Monitoring Using Real-Time Wearable Devices}
\label{sec:intro1}
Wearable devices have received increasing interest from hospitals as they allow clinical decisions to be based on continuous measurements instead of a few measurements acquired by trained clinical staff~\cite{LDSZZSPCDMS:17,KRWJ:19,RWTR:20,PKTKTBG:20,QRGBARKT:20,dorr_iphone_2021}. 
During hospitalization, health data is usually obtained manually at single points in time by health care providers, who then enter the data into the hospital's clinical information system at a later point in time if no automated infrastructure is present. This procedure is time consuming for the care team and often results in sparse data with inaccurate timestamps. Wearable devices offer a convenient, comfortable and affordable means of acquiring patient health data on an automated real-time 24/7 basis, both during hospitalization and beyond.

Remote or unattended patient monitoring can be especially beneficial in the context of highly infectious diseases such as COVID-19~\cite{SDHHKWVD:20,QRGBARKT:20,MWMBBBACH:20} 
where traffic between infected patients needs to be minimized and at the same time frequent documentation of vital signs is required. Acquiring disease relevant vital signs using wearable sensors enables real-time remote monitoring of patients’ health and at the same time reduces the number of contacts with infected patients. Moreover, the use of wearables is a simple and an inexpensive approach to meeting any increased short-term demand for patient monitoring during pandemics. High temporal resolution, wearable health data has the potential to provide fast real-time notifications of alarming patient conditions. It is important to understand that the simultaneous documentation of multiple vital signs produces a vast amount of information that cannot be reviewed by the care team alone with the resources available to them. On an intensive care unit, with a comparable density of data, a physician is responsible for only a few patients. However, in a regular ward a physician takes care of numerous patients at a time. Therefore, the data has to be processed and presented in a way that can be rapidly and reliably interpreted.

Monitoring of vital signs is a clinical routine with known challenges~\cite{AFGR:12}. In contrast, real-time monitoring of patients using wearables is not yet established and adds a set of unique challenges to medical time series analysis and visualization. Aigner et al.~\cite{AFGR:12} presented six challenges for visual analytics in healthcare that are relevant to this work: \textit{(1) scale and complexity of time-oriented data, (2) intertwining patient condition with treatment processes, (3) scalable analysis from single patients to cohorts, (4) data quality and uncertainty, (5) interaction, user interfaces, and the role of users, and (6) evaluation}. In addition to addressing challenges (1), (4) and (6), this study focused on the communication of noteworthy patient health patterns recognized through visual analysis from continuous wearable data. The fluctuation of data quality due to compliance, battery charging, improper use or technical issues is especially relevant for wearable health data. Further visualization challenges of medical time series arise from the fact that vital sign values, such as heart rate, are patient specific, while others such as temperature, are physiologically similar for all patients. Additionally, vital signs such as the heart rate are bi-directional in nature, i.e. values that are too high and too low are both alarming, while others, such as arterial oxygen saturation, are one-directional, i.e. the lower the value, the greater the cause for alarm.

There are other specific challenges for wearable data. Only a few manufacturers offer real-time wearable devices that are certified medical products~\cite{biofourmis,e4,ActiGraph}. 
In addition, most manufacturers certify only a subset of device parameters, such as heart rate or arterial oxygen saturation~\cite{biofourmis}, which have been validated on small and mostly healthy populations. 
Furthermore, device manufacturers calculate derived signals based upon proprietary algorithms that are not medically certified, e.g. respiration rate or core body temperature. 
It is important that the assessment of the patient condition is not compromised by unreliable sensory data. Therefore, it is particularly important to consider data quality. 
To this end, some device manufacturers, e.g.~\cite{biofourmis}, offer data quality estimates of selected vital signs that can be used to clean the data. 
Alternatively, quality indices could potentially be computed from raw data~\cite{KKAD:12,GKLZK:20}. 

\subsection{Intuitive Visualization of Wearable Health Data}

As humans are skilled at recognizing visual patterns~\cite{Shneiderman:96,TFCS:11}, visualizations can be used to quickly spot missing data or unusual health patterns during patient monitoring and treatment optimizations~\cite{AFGR:12}. Effective means of data exploration save busy health care providers time by effectively spotting relevant changes in patient health trajectories. In return, this leads to a faster understanding of complex information and potentially reduces the number of human mistakes. 
However, because of the high rates at which wearable health data is recorded (1--50Hz), it is helpful to condense the information contained in the data, such as by reducing the temporal resolution or by reducing the number of vital signals displayed. 

There are numerous visualization methods for time-oriented data available, especially for linear vs. cyclic time and single time points vs. time intervals~\cite{AMMST:08,AMST:11}. 
Classic time-oriented visualizations are line plots that display the position of variable values along the time axis. Bar charts with upward and downward bars are used to depict relative changes in magnitude, where upward bars represent values above and downward bars represent values below a typical value. Additional variables can be encoded in visualizations using shapes, textures or colors. Advanced time-oriented information visualizations methods~\cite{AMMST:08,AMST:11} include: SpiralGraphs, which allow periodic patterns to be visualized; ThemeRiver graphics, which are used to display thematic changes in a given variable collection; and Midgaards, which allow representations to be resized to varying levels of detail. However, these advanced methods were not considered since, based on initial discussions with the clinical staff, they were deemed to be too complex.

To visualize wearable health data over the course of a day, Frink et al.~\cite{FGM:17} implemented hourly circular pie charts for health data recorded from mobile applications, resulting in a circular form of data visualization similar to SpiralGraphs. Another wearable health visualization study~\cite{AFMJ:17b} developed a visual dashboard depicting aggregated patient activity as hours per days (and minutes per hours) using color-coded upward pointing bar charts. We visualized hourly wearable data, as this is a natural level of granularity that quickly provides an overview for visual pattern detection. However, unlike previous studies, we evaluated perceptually motivated heat maps and color-coded upward and downward pointing bar charts.

Specifically, our contributions are:
\begin{itemize}
    \item	Evaluation and improvement of compact and easy to read visualizations to communicate noteworthy clinically relevant patterns from wearable health data via visualizations.
    \item Analysis and specification of the typical challenges of using wearable health data in clinical settings.
    \item Exploration of a wearable health dataset collected from 84 hospitalized COVID-19 patients.
\end{itemize}


\section{Application Domain: Patient Monitoring Using Wearables}
\subsection{Clinical Use Case}
\label{sec:use_cases}

In this study, the primary use case was the communication of prominent patient health patterns between clinical staff and data scientists using visualizations. Clinical staff frequently ask for support in developing models or algorithms to automate the detection of abnormal health patterns or to derive bio-markers. In order to develop a robust model for this use case, it is important to work with high quality data that captures relevant clinical events. Clinical staff need to explain to data scientists what typical or atypical data readings may look like. Moreover, in the course of the development of a new (predictive) model, it is helpful to visualize algorithm results in order to discuss and explain computer decisions, for example when using explainable artificial intelligence (XAI) approaches~\cite{Barredo:20}.  

Other use cases include real-time ward monitoring with technical trouble shooting or real-time detection of patients in critical conditions. The latter use case requires health care providers to be alarmed when a critical situation is imminent or is already occurring. In particular, continuously increasing or decreasing trends in signals from a patient should raise an alarm. For example, a heart rate that is continuously increasing might still be within a normal range, but may already indicate a cardiovascular problem in the patient. If a change in a vital parameter is automatically detected over time, interventions can be applied sooner. The results from the primary use case are expected to be transferable to the other mentioned uses cases.

\subsection{Data}
\label{sec:data}

Wearable health data was acquired from 84 patients that were hospitalized due to the detection of severe acute respiratory syndrome coronavirus 2 (SARS-CoV-2) in the COVID-19 ward of the University Hospital Basel, Switzerland, from March to May 2020. The dataset contains 30 female and 54 male patients (mean age: $60 \pm 17$ years, range: 17-97 years). On average, patients were admitted for $6.6 \pm 4.7$ (range: 0.1-23.1) days and data was recorded for $5.6 \pm 5.4$ (range: 0.0-26.1) days (see Fig.~\ref{fig:data_days}). No valid data could be recorded for five of the patients (four females, one male). 19 patients wore the device for less than two days; 24 patients wore the device for between two and five days; 26 patients wore  the device for between five and ten days; and ten patients wore the device for between ten and 30 days.

During heir hospitalization, the patients wore a medically certified Everion 3.06 wearables device from Biofourmis (formerly Biovotion)~\cite{biofourmis,everion:17}. The health data was acquired using the Everion mode that recorded 22 vital signs plus seven quality signals for selected vital signs at a sampling rate of 1Hz. Of the 22 signals recorded, the clinical staff considered the following signals to be most relevant for the COVID-19 patients: heart rate (HR), HR variability (HRV), respiration rate (RR), oxygen saturation (SPO2), and core body temperature (Temp). HRV is derived from the HR signal and represents the variations in time between heart beats. The Everion device calculates the root mean square of successive time differences (RMSSD) between heart beats measured in milliseconds. The RR is usually estimated based on the derived HRV and the core body temperature is usually calculated based on the skin temperature and other signals (see e.g.~\cite{BTCMKCRRCJA:13,EMKBRA:18,SDHHKWVD:20}). The RR and core body temperature estimation methods used by the Everion are proprietary. Acquisition of the HR and SPO2 signals has been medically certified for the Everion device. 

The clinical staff sporadically tested the reliability of some signals themselves and reported the following subjective assessment: the HR was a reliable and robust parameter; the RR was mostly reliable but sometimes incorrect; the SPO2 values were plausible, but only rarely recorded; the core temperature was error-prone and not reliable for medical purposes. Specifically, it was observed that the temperature signal increased when a patient put their arm below the blanket. Despite these observations, we still used the core temperature for visual exploration because of its medical relevance. Nevertheless, this parameter must be considered with caution.

Vital signs from a patient with a COVID-19 infection can be expected to change as follows  (see also~\cite{COESB:20,SDHHKWVD:20}):

\begin{itemize}
    \item HR: decreases
    \item HRV: decreases
    \item SPO2: decreases with reduced lung function, indicating a worsening patient condition
    \item RR: increases with severity of COVID-19 symptoms, reflecting a worsening condition of the lungs 
    \item Temp: elevated
\end{itemize}

Note: The HR was observed to decrease during COVID-19 infections~\cite{COESB:20}, in contrast to a known HR increase for viral diseases in general~\cite{SDHHKWVD:20}.

\subsection{Typical Challenges of Wearable Health Data}
\label{sec:data_problems}

Wearable data quality is typically highly volatile (Sec.~\ref{sec:c_missing}--\ref{sec:c_quality_inspection}). According to Gschwandtner et al.~\cite{GGAM:12}, who developed a taxonomy for ``dirty'' time-oriented data quality problems, wearable health data is \textit{single-source rastered interval} data with a beginning and end, but with data gaps in between. According to their taxonomy, wearable health data contains missing data (values or tuples) and implausible values (unexpected high/low values). Furthermore, it can be wrong or unusable~\cite{KCHKL:03}. Challenges specific to the visualization of medical time series are outlined in Sec.~\ref{sec:c_sampling_rate}--\ref{sec:c_spot_patterns}.

\subsubsection{Challenge: Missing Data}
\label{sec:c_missing}
Data may often be missing, because patients do not always wearing the device properly or may remove the device temporarily. The battery duration of the devices depends both on their capacity and on the acquisition mode (raw data vs. aggregated data). For the COVID-19 patients, the wearable devices were typically charged once a day in the morning hours after waking.

\subsubsection{Challenge: Implausible Values}
\label{sec:c_implausible}
Another challenge is the heterogeneous data quality of varying signals. Some signals, such as HR, are quite reliable, while others, such as core temperature, are derived estimates and need to be treated with caution (see Sec.~\ref{sec:data_problems}. 
For example, the device might still be recording data such as ambient temperature, even when it has been removed from the patient (Sec.~\ref{sec:c_missing}. As can be noted from the admitted vs. recorded data day numbers in Sec.~\ref{sec:data}, this challenge that is specific to wearable data was also observed in our data (the average of recorded data days was higher than the average of admission days); specifically, for 8 patients.

\subsubsection{Challenge: Careful Quality Inspection Required}
\label{sec:c_quality_inspection}
In order to reduce poor-quality data, post-hoc individual data point quality analysis and cleansing is essential for clinical purposes. In certain situations, it is however difficult to estimate whether a certain data outlier is an artifact or a dangerous patient situation.

\subsubsection{Challenge: Sampling Rate}
\label{sec:c_sampling_rate}
The typically high sampling rate of acquired information (our data: 1--50Hz) makes it cumbersome to scan through the data to discover relevant vital parameter changes. Therefore, complexity with respect to temporal resolution and the number of displayed variables needs to be reduced in order to make data comprehensible.

\subsubsection{Challenge: Individual Data Ranges}
\label{sec:c_data_ranges}
Data ranges of certain vital signs, such as HR or RR, are highly individual, while other vital signs, such as Temp,  are physiologically similar for all patients. Therefore, upper and lower value bounds need to be individually set for every patient – either calculated or by clinical personnel.

\subsubsection{Challenge: Consistent Color-Coding}
\label{sec:c_color_coding}
The challenge of color-coding the various vital signs in a consistent way is directly connected to the challenge in Sec.~\ref{sec:c_data_ranges}. Certain signals, such as the HR, are divergent in nature, i.e. a patient has a baseline value range where values above or below have a different medical interpretation. Other signals, such as the SPO2, are sequential in nature  where lower data values correspond to a decrease in condition or vice versa, such as for HRV. This quickly leads to a large spectrum of colors and meanings for different signals.

\subsubsection{Challenge: Spot Irregular Patterns}
\label{sec:c_spot_patterns}
Clinical staff need to be able to spot irregular health conditions (patterns) at a single glance.

\section{Methods}

\begin{figure*}[htb]
\includegraphics[width=1\linewidth]{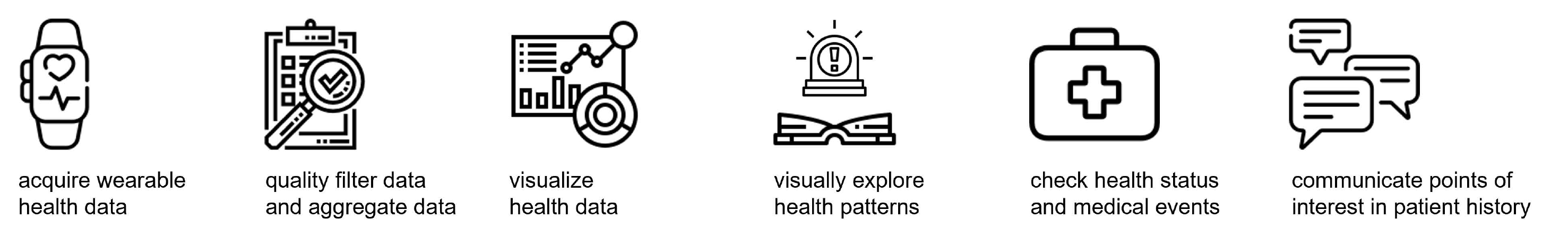}
 \centering
  \caption{\label{fig:overview}Wearable health data pipeline for the use case of communicating relevant health patterns between clinical staff and data scientists through visualizations.}
\end{figure*}

Our method workflow is illustrated in Fig.~\ref{fig:overview}, which will be explained in more detail in this section starting with the acquired wearable health data as described in Sec.~\ref{sec:data}.

\subsection{Wearable Health Data Quality Filtering}

To address the challenge that wearable data may be available during periods where the device was not worn (during admission or after discharge, Sec.~\ref{sec:c_implausible}), we filtered the data according to the heart beat. In other words, we set all of the signal values to NAN if no HR was detected at a given point in time. We chose this procedure, since the HR is considered to be a highly reliable vital parameter that can be acquired from wearable devices (Sec.~\ref{sec:data}). In cases where no HR is recorded, but other data is recorded, it is likely that the device is switched on but not properly attached to the patient (Sec.~\ref{sec:data_problems};\ref{sec:c_implausible}). 

Since careful quality inspection is required for wearable health data (Sec.~\ref{sec:c_quality_inspection}), we cropped the available data by using the admission dates and quality filtered the data. According to the Everion device documentation~\cite{biofourmis,biofourmis:20}, vital signs with a quality of at least 50 percent can be considered medically reliable. The calculation methods for the quality parameters of the Everion device are proprietary. For simplicity, we filtered the data to keep only signals with 50--100 percent data quality. More advanced methods to replace missing-values or to filter~\cite{KKAD:12,GKLZK:20} the data were not considered at this point. One reason was that the data from the wearables was acquired at a frequency of 1Hz, while the visualizations were generated using hourly aggregated data.

\begin{figure}[htb]
  \centering
  \includegraphics[width=0.6\linewidth]{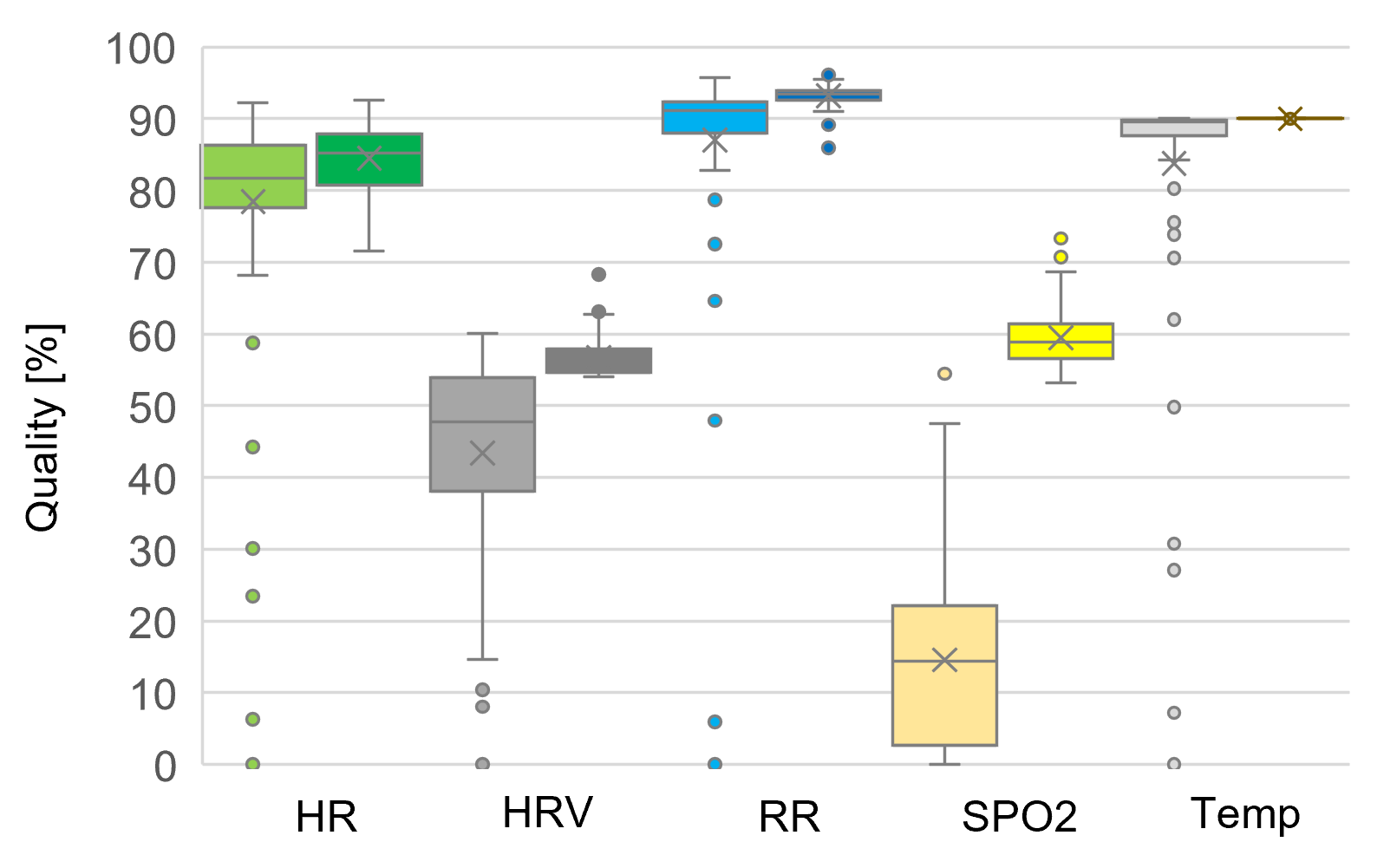}
  \caption{\label{fig:sig_quality}
           Average quality of the selected vital signs for all 84 patients before (left) and after (right) quality filtering.}
\end{figure}

\begin{figure*}[htb]
  \centering
  \includegraphics[width=1\linewidth]{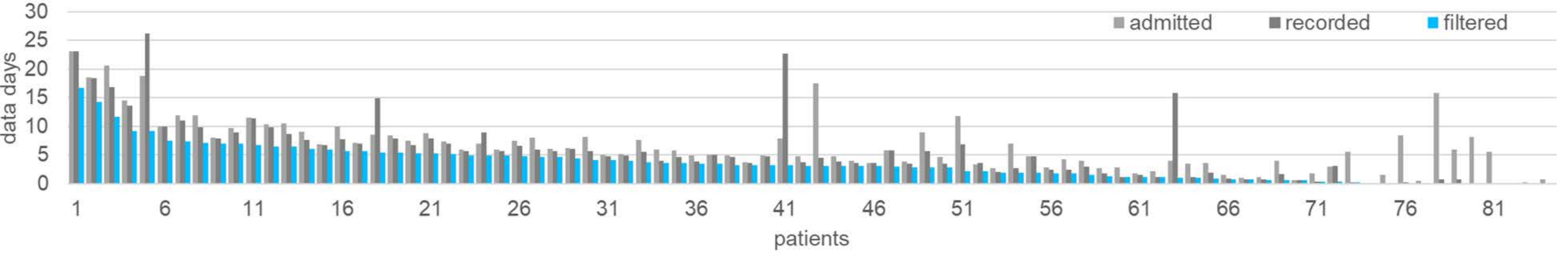}
  \caption{\label{fig:data_days}
           Admitted patient days (light gray); recorded wearable health data days (dark gray); useful quality filtered data days (blue), i.e. non-missing values, for all 84 COVID-19 patients.}
\end{figure*}

The data quality before and after quality filtering is shown in Fig.~\ref{fig:sig_quality}. Useful filtered data was available on average for 84 hours or $3.5 \pm 3.2$ days (range: 0.0-16.8) - excluding time points with missing values. Out of the recorded 5.6 data days, on average roughly two third of the data was usable. As depicted in Fig.~\ref{fig:data_days}, useful quality-filtered data was generated for up to two days from 27 patients, for two to five days from 30 patients, for five to ten days from 19 patients, and for ten to 17 days from three patients. No data was recorded from five patients.

\subsection{Data Aggregation}

The clinical staff initially requested only a few values per day (challenge Sec.~\ref{sec:c_sampling_rate}); however, we concluded that hourly values were beneficial since they depict daytime and nighttime patterns well, e.g., circadian rhythms (similar to~\cite{FGM:17,AFMJ:17b}). The chosen hourly time scale can be naturally read by humans as it represents the 24 hours clock cycle. We aggregated the acquired vital sign values from seconds to hourly averages using the arithmetic mean of all available non-missing values. In general, the arithmetic mean works well if data is normally distributed and does not suffer from outliers. From our raw data we did not observe any significant outliers and hence we chose this method. It was not our focus to compare alternatives for  data aggregation. However, this could be elaborated in more detail in the future.
This hourly scale reduces the initial high raw sampling density and smooths out short-term fluctuations. Longer periods of missing data or data that was discarded due to unreliability can easily be spotted in the visualizations. 

\subsection{Visualization Methods}
\label{sec:vis}

\begin{figure}[htb]
  \centering
  \includegraphics[width=0.6\linewidth]{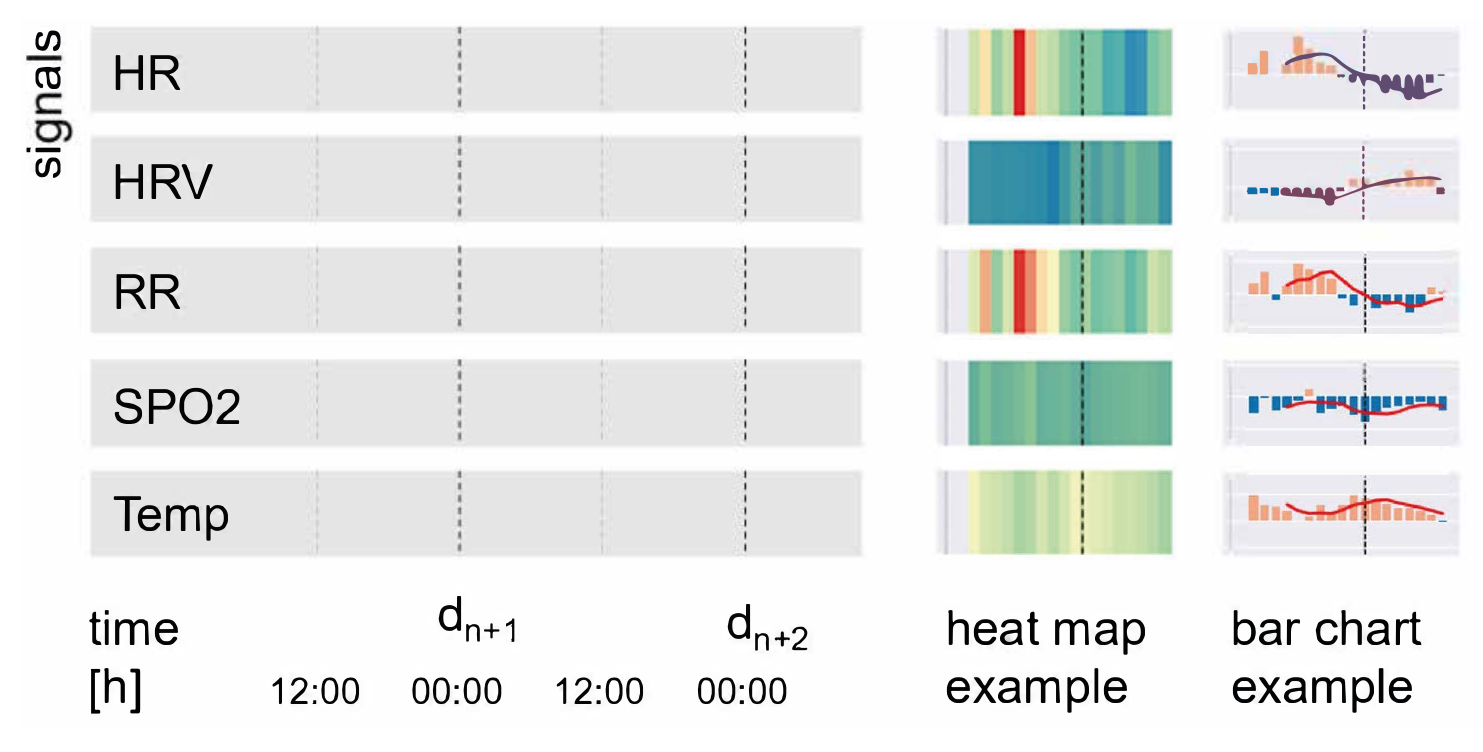}
  \caption{\label{fig:concept}
           Left: Customized plotting layout for the visual health data. Right: heat map and bar chart example visualizations. }
\end{figure}

According to Aigner et al.~\cite{AMST:11} a time-oriented visualization include: what is presented (time and data); why is it presented (user); and how is it presented (visual representation). In our visualizations, we displayed post-hoc wearable vital signals over the full course of patient hospitalizations (admission) on the COVID-19 ward of the University Hospital Basel (Sec.~\ref{sec:data}). Moreover, despite the suggestion by Halford et al.~\cite{HBMB:05} to use four variables at most, we decided to use five vital signals (HR, HRV, RR, SPO2 and Temp), since they were considered clinically relevant (Sec.~\ref{sec:data}). 
The users are data scientists, who initially inspected the data for quality purposes, as well as clinical staff, who evaluated the data using matching data from the clinical information system. Specifically, we developed heat maps and bar charts together with a  custom-designed plotting layout~\cite{SRWJ:21} to make the data intuitively readable (challenge Sec.~\ref{sec:c_spot_patterns}), as illustrated in Fig.~\ref{fig:concept}.

The visual structure took into account perceptual mechanisms that allow viewers to orient themselves within the visualizations and was designed according to Gestalt principles: \textit{similarity} was used in the design of five parallel bands of data for the major vital signals that are structured analogously; \textit{proximity} was used between the slightly separated major signals to make them appear as one group of data along the same timeline; \textit{connection} was used along the time axis where empty data is padded and made clearly visible along one band; \textit{continuity} and \textit{closure} were used to mark general timeline points such as midnight (black dashed lines) and noon (grey dashed lines) on individual signal bands in order to provide the user with an efficient orientation point. In order to draw the viewer's attention to specific areas, visual   highlighting mechanisms such as color or size were used. The ordering was performed along the x-axis according to time and along the y-axis according to alphabetic order of the signal abbreviations. Selective information about medical events was manually added to these prototype visualizations by the data scientists, which consisted of text noteworthy with observations. 

As shown in Fig.~\ref{fig:concept}, we indicated major tick labels for days and minor tick labels for hours along the time axis. In our developed framework~\cite{SRWJ:21}, the exact date is displayed in the format YYYY-MM-DD. However, for this publication we use days $d1, \dots, d_N$ to remove sensitive patient information. The minor tick labels are formatted as HH:MM starting at midnight, i.e. 00:00. Depending on the length of the displayed time interval $t$, we labeled: every hour for $t < 24 h$; every sixth hour for $1d \leq t < 6d$; every twelfth hour for $6d \leq t < 10d$; and every twenty-fourth hour for $t \geq 10d$.

\subsubsection{Heat Maps Using Clinical Color-Coding}
\label{sec:heat_maps}

\begin{figure}[htb]
  \centering
  \includegraphics[width=0.48\linewidth]{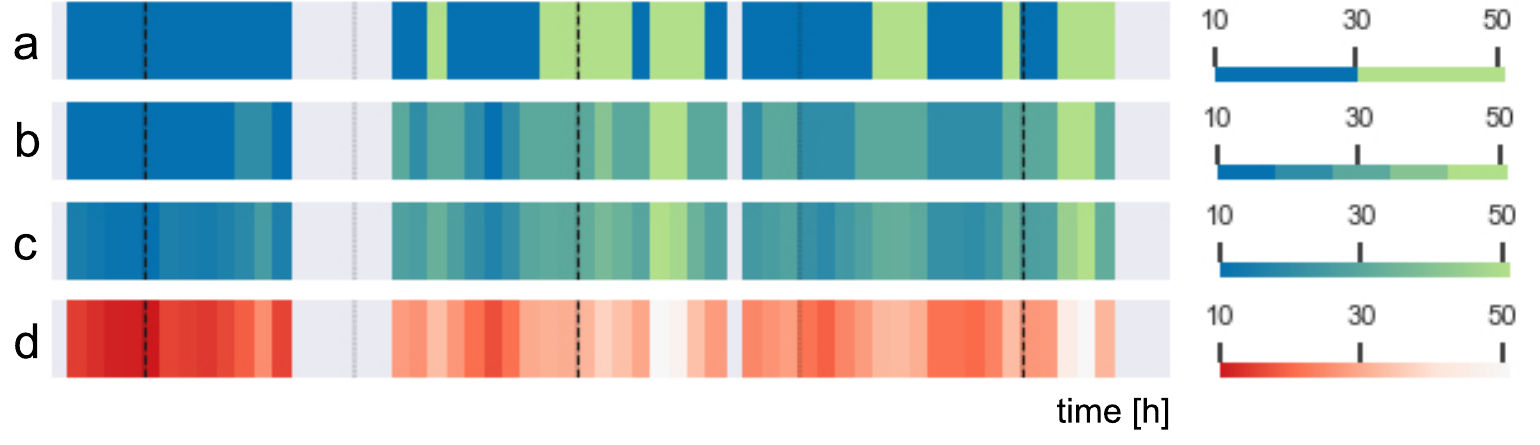}
  \caption{\label{fig:hrv}
           Sequential color scheme variations, such as for HRV.
}
\end{figure}

\begin{figure}[htb]
  \centering
  \includegraphics[width=0.48\linewidth]{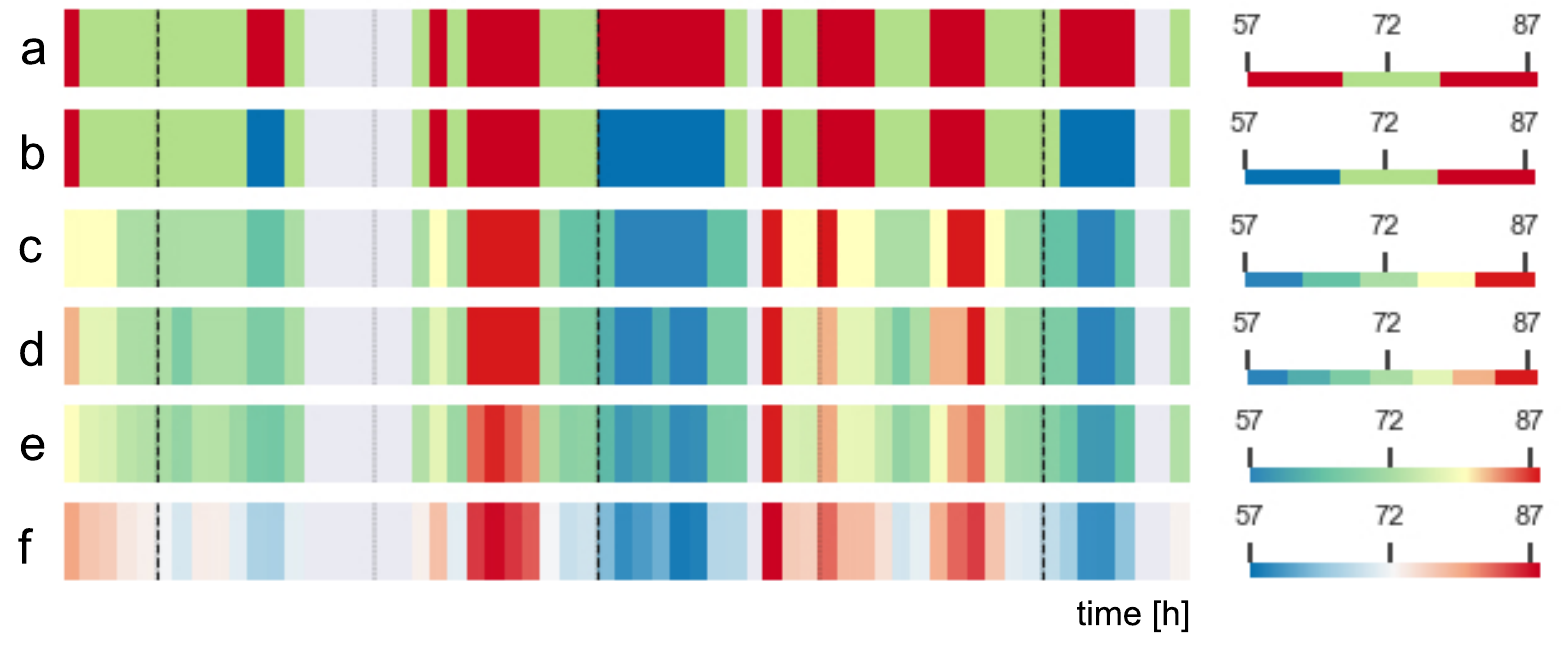}
  \caption{\label{fig:hr}
           Diverging color scheme variations, such as for HR.
}
\end{figure}

Heat maps encode the intensity of a value using color. Clinical color-coding conventions can depend on the use case and context. Generally, red is used to depict an alarming situation and for values that are too high. Blue is used for too values that are too low. However, for signals such as SPO2, values that are too low are alarming and hence should be red and blue at the same time. White or green are used for data values in acceptable ranges. Yellow (orange) is used for ambiguous situations that are not yet alarming, but require investigation. In order to evaluate this color-coding challenge (Sec.~\ref{sec:c_color_coding}) for the visualization of the hourly wearable health data, we tested various color schemes. Sequential color schemes are appropriate for data that is ordered from low to high values~\cite{Brewer:94} such as HRV and SPO2. Diverging color schemes are well suited for data where most of the data is in the mid-range, but contains extreme values near both ends of the data range~\cite{Brewer:94} such as HR, RR and Temp. 
Sequential color schemes are usually constructed such that light colors reflect low data values and dark colors depict high data values~\cite{Brewer:94}, e.g., HRV in Fig.~\ref{fig:hrv}. However, for SPO2 we figured that the light colors should reflect high values and dark colors should reflect low values: The lower the SPO2 value, the more alarming the patient's situation. For diverging color schemes, the low and high extremes are highlighted with dark values while the central data range is represented using light colors~\cite{Brewer:94}, such as for HR in Fig.~\ref{fig:hr}.

We selected the initial color schemes from Colorbrewer~\cite{BHSWH:20}: white-red (WR) (Fig.~\ref{fig:hrv}d) or yellow-red (YR) for sequential schemes and red-white-blue (RWB) (Fig.~\ref{fig:hr}f) or red-yellow-blue (RYB) for diverging schemes. After analysing the color-coding using a survey (Sec.~\ref{sec:appendix}), we adapted the Colorbrewer schemes to use green for the central normal range values (Fig.~\ref{fig:hrv}a--c and Fig.~\ref{fig:hr}a--e) and yellow for ambiguous values above the baseline for diverging schemes (Fig.~\ref{fig:hr}c--e). We did not encode ambiguous values in yellow below the baseline, because it was considered confusing. The decision to use these colors was taken during a consensus meeting after the written survey had been completed. In addition, we tested discrete (Fig.~\ref{fig:hrv}a and Fig.~\ref{fig:hr}a,b) vs. continuous (Fig.~\ref{fig:hrv}b--d and Fig.~\ref{fig:hr}c--f) color schemes. For simple use cases such as ward monitoring where clinical staff only have to determine whether the recorded data is in a normal range or not, green-blue (GB) can be used for sequential schemes (Fig.~\ref{fig:hrv}a) and red-green-red (RGR) (Fig.~\ref{fig:hr}a) or red-green-blue (RGB) (Fig.~\ref{fig:hr}b) for diverging schemes.

\subsubsection{Bar Charts}

Similar to Aigner et al.~\cite{AMMST:08}, we used color-coded bar charts with upward and downward bars to depict signal trends, patterns and relative changes in the patient's health condition (challenges Sec.~\ref{sec:c_color_coding}--\ref{sec:c_spot_patterns}). The baseline to separate upward and downward pointing bars was calculated for each signal as arithmetic mean over the full time period of the patient hospitalization (admission) on the COVID-19 ward. We chose this approach in order to visualize relative changes within a selected time frame independent of the current condition of the patient. Based on the evaluation of the color conventions by the clinical staff (Sec.~\ref{sec:heat_maps}), we used a light red-like color for values above the baseline and blue for values below the baseline with varying hues to render the two colors even more distinct. Additionally, the moving average over the previous four hours was plotted as a red line (no color-coding applied) to display a smoothed trend.

\subsubsection{Individual Patient Data Ranges}
Striking a balance between global and individual value ranges is a challenge~\cite{AMST:11}, because the value ranges for many signals such as HR, HRV or RR are often patient-specific. In contrast, value ranges for signals such as SPO2 or Temp can be expected to be in a similar range for all patients. In order to address this challenge (Sec.~\ref{sec:c_color_coding}), the signal value ranges and the baselines for the bar charts were defined in the following way: By default, no value ranges were set and the minimum, maximum and mean values over the full displayed time period were calculated individually for each patient. However, these ranges can be configured for each signal, e.g., when clinical staff decide to modify them manually.

\subsubsection{Evaluation of Visualizations}
\label{sec:eval}

The heat map visualizations were evaluated using a written survey that was answered individually in by three persons from hospital (two medical doctors and one clinical project manager) and seven health data science researchers, as detailed in the Sec.~\ref{sec:appendix}. The individual survey answers were discussed in a consensus meeting with two clinical staff and three health data scientists. The written survey contained eight questions related to the chart features, the color-coding, the usability and suggestions for improvement (see challenges in Sec.~\ref{sec:data_problems}). Two extra questions related to the signal value ranges and  clinical relevance were only posed to the persons working in the hospital. The subjective usability of the heat maps was assessed via a simplified version of the system usability scale (SUS)~\cite{Brooke:96,SUS:20}. Evaluation of the bar charts was performed during a review meeting with two clinical staff and three health data scientists.

Using the developed framework~\cite{SRWJ:21}, we auto-generated heat maps and bar charts for all of the COVID-19 patients. Two data scientists selected 20 interesting cases where typical and noteworthy patient health condition trajectories were discovered. These 20 patients were then inspected by clinical staff who matched the visualizations with medical events recorded in the clinical information system. During this visual examination of the wearable health data, one typical and two noteworthy COVID-19 patients were selected.

\section{Results}

\begin{figure*}[htb]
  \centering
  \includegraphics[width=1\linewidth]{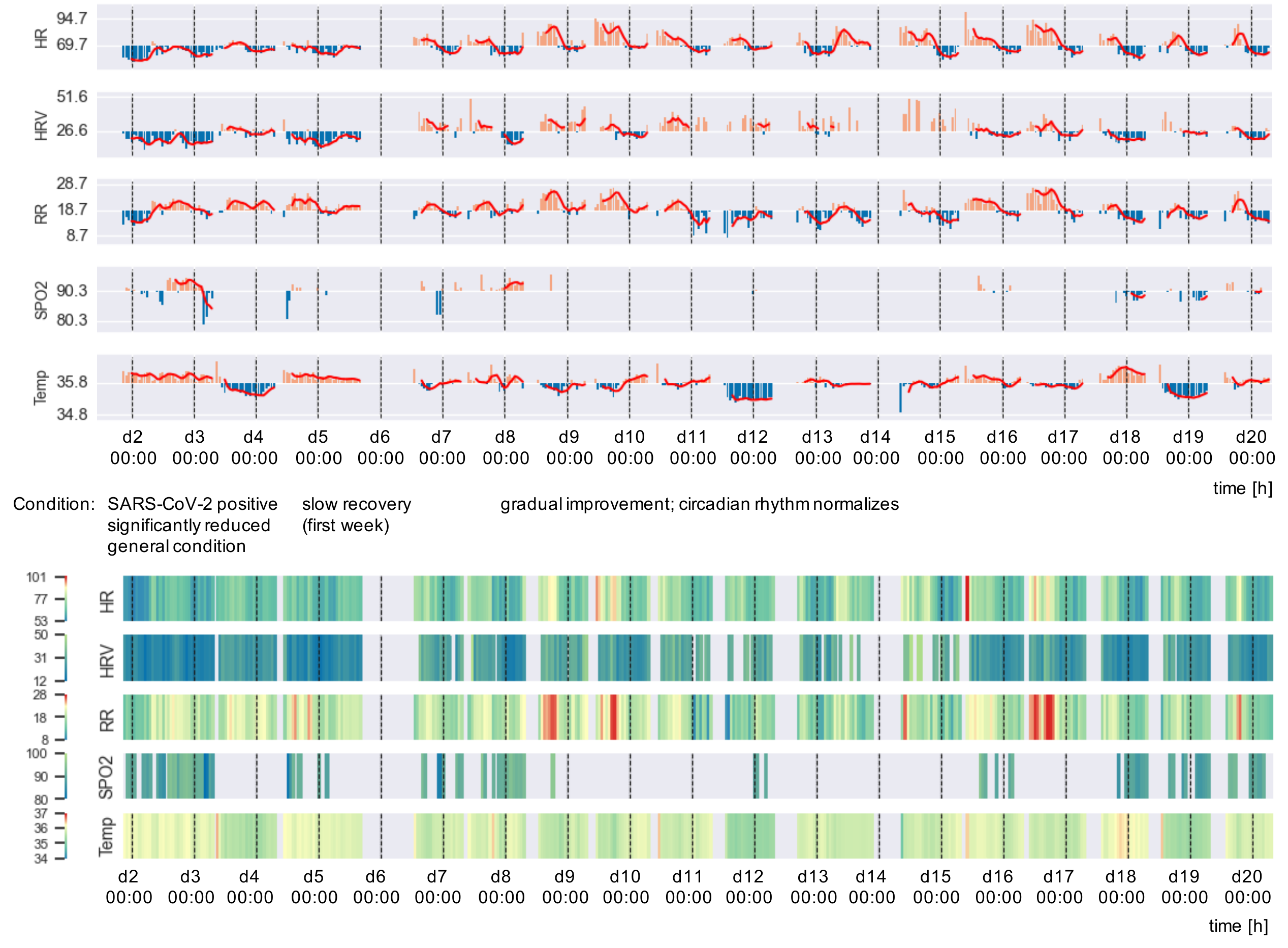}
  \caption{\label{fig:patient-normal}
           Visualizations of vital signs from a typical COVID-19 patient with annotations of the health condition. Wearable data was recorded from day 2 to day 20. The bar charts clearly show pattern changes (HR, HRV and RR) throughout the recovery. The heat maps particularly highlight critical values (in red).
}
\end{figure*}
 
\begin{figure*}[htb]
  \centering
  \includegraphics[width=1\linewidth]{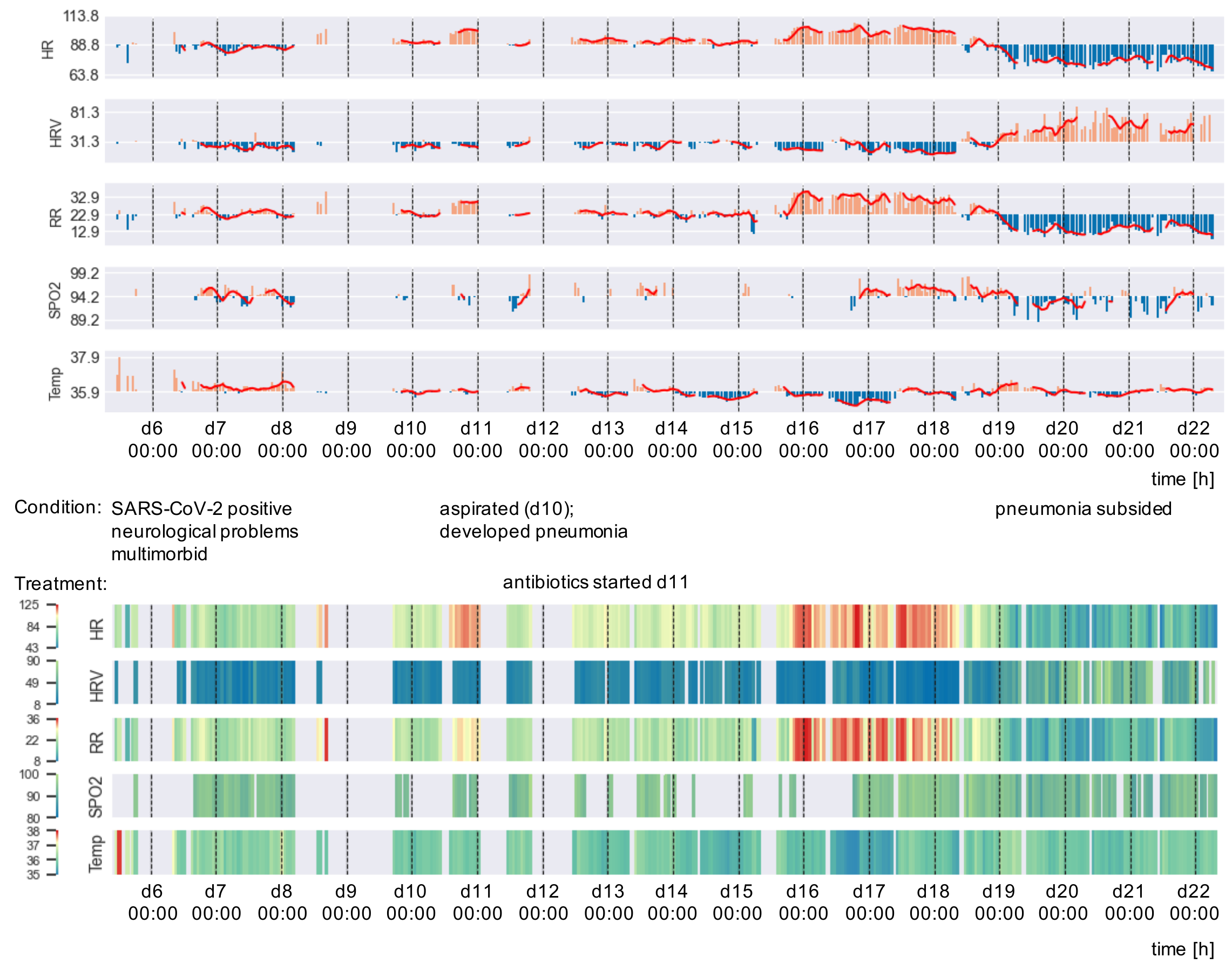}
  \caption{\label{fig:patient-complications}
            Visualizations of vital signs from a COVID-19 patient who received medication due to the annotated complications. Wearable data was recorded from day 6 to day 22. The bar charts clearly show pattern changes (HR, HRV and RR) throughout the recovery. The heat maps  highlight critical values (in red).
}
\end{figure*}

\begin{figure*}[htb]
  \centering
  \includegraphics[width=1\linewidth]{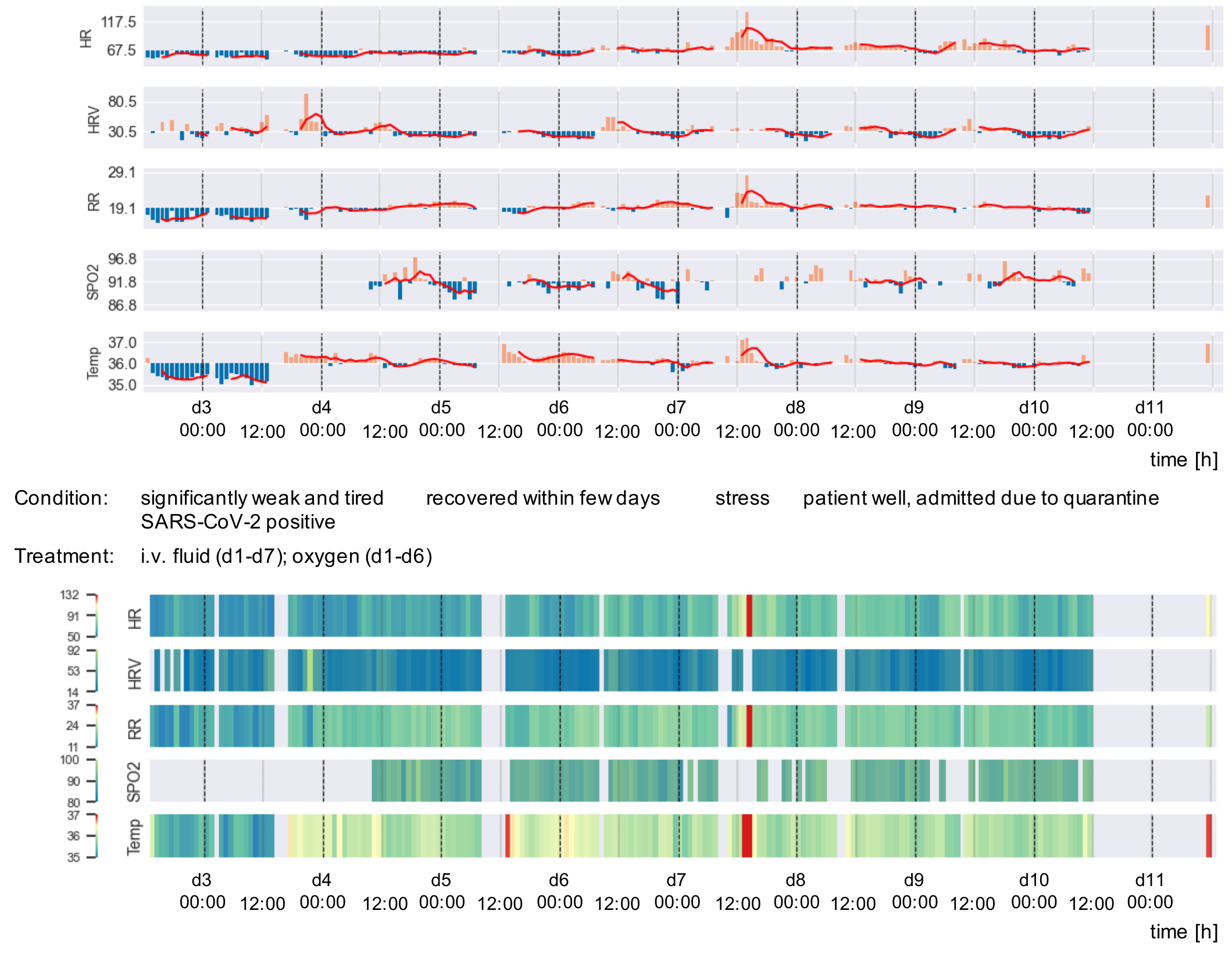}
  \caption{\label{fig:patient-stressed}
           Visualizations of vital signs from a COVID-19 patient experiencing a stressful short-term situation.
           The patient received medication due their initial weak condition, as annotated. Wearable data was recorded from day 3 to day 11. From left to right, a general trend  from lower to higher vital sign values (HR, RR and Temp) is visible. There is some evidence of periodic HR patterns visible in the bar charts. The stress situation is clearly highlighted in red in the heat maps and as major peaks in the bar charts.
}
\end{figure*}

\subsection{Evaluation Results}

\subsubsection{Evaluation of Color-Coding and Heat Maps}
The results of the evaluation of the color-coding (challenge Sec.~\ref{sec:c_color_coding}) and the heat maps are based on the visualization survey described in Sec.~\ref{sec:eval} and detailed in Sec.~\ref{sec:appendix}. In general, the heat maps were considered to be intuitive by providing an overview (challenge Sec.~\ref{sec:c_spot_patterns}) of the patient health condition evolution at a glance, showing trends and critical values. One data scientist pointed out that such visualizations are a useful tool for communicating with clinical staff and discovering medical events -- our primary use case. 
All three clinical staff clearly stated that the heat map visualizations were useful for clinical purposes. 
With respect to the SUS~\cite{Brooke:96,SUS:20} evaluation, seven individuals (six data scientists, one person from the hospital) stated that they thought that most people would quickly learn how to use these visualizations. Four individuals reported that they would use the visualizations frequently and three individuals felt confident in using the visualizations. One individual from the hospital reported that the visualizations were unnecessarily complex and cumbersome. The two other individuals from the hospital stated that teaching and technical support is needed for clinical use by care teams -- not our primary use case. 

Some ambiguity in the color schemes was detected and resolved (Sec.~\ref{sec:heat_maps} and challenge  Sec.~\ref{sec:c_color_coding}). 
Five individuals voted for RWB, four individuals voted for RYB and two persons suggested using green for normal range values and yellow for ambiguous values. In the end, we decided to use Colorbrewer adapted color schemes as described in Sec.~\ref{sec:heat_maps}. Five individuals voted for discrete and continuous color schemes. Since for our primary use case the communication of patterns is important (challenge Sec.~\ref{sec:c_spot_patterns}), we decided to adopt a continuous color schemes with GB for sequential schemes (Fig.~\ref{fig:hr}c) and RYGB for diverging schemes (Fig.~\ref{fig:hrv}).

\subsubsection{Evaluation of Visualization Plotting Layout}
The visualization evaluation led to improvements in the layout and features of the plotted data (challenge Sec.~\ref{sec:c_spot_patterns}), which were integrated into the final visualization concept, as shown in Fig.~\ref{fig:concept} and described in Sec.~\ref{sec:vis}. 
Specifically, the suggested improvements were: major ticks for days and minor ticks for hours; time denoted as HH:MM instead of HH; signal separation using spacing; a gray background and green for normal range values.

\subsubsection{Evaluation of Bar Charts vs. Heat Maps}
\label{sec:chart_type_eval}
All three persons working at the hospital felt that the bar charts clearly depict the evolution of the patient's health condition, regular vs. irregular patterns as well as circadian rhythms. The heat maps, however, were found to be helpful in identifying extreme values that might reflect critical patient situations. The clinical staff found both visualization types easy to use  (challenges Sec.~\ref{sec:c_color_coding}--\ref{sec:c_spot_patterns}), albeit with a preference for the bar charts. 

\subsection{Analysis of Implausible Values}
In terms of the challenge outlined in Sec.~\ref{sec:c_implausible}, we found that the average core body temperature after quality filtering of our dataset was around 36\textdegree C, which seemed too low for patients suffering from a disease that caused fever, especially since the expected value for healthy people is 37.5\textdegree C (see subjective observations in Sec.~\ref{sec:data}).

\subsection{Visual Exploration of COVID-19 Patients' Conditions}

In this section, we examine three COVID-19 patients: one typical COVID-19 patient history; one patient history illustrating complications; and one patient history that includes a short-time period of stress. For all patients, we depicted heat maps as well as bar charts to illustrate the value of each visualization type, as described in Sec.~\ref{sec:chart_type_eval}.

\subsubsection{Normal COVID-19 Case}

The patient history in Fig.~\ref{fig:patient-normal} shows a typical disease development of COVID-19. The SARS-CoV-2 positive patient was a 75-year-old man who was admitted to the COVID-19 ward as a result of his significantly reduced general condition. After a rather slow recovery during the first week, his condition gradually improved from day 8 on (HR during daytime increased, HRV recovered). His HR was relatively low in the first week (typical for COVID-19~\cite{COESB:20}). Then, possibly due to more activity, his HR increased over the course of his continued hospitalization. The circadian rhythm is clearly visible in the bar chart visualizations (dashed black lines represent midnight). It can be seen that the circadian rhythm in the HR, HRV and RR was initially quite irregular, but this normalized during the last three days of hospitalization. The low HRV at the beginning of the stay, which increased during recovery, is a known sign of a severe infection~\cite{HGDHWHC:20}. Between days 16 and 17, prominent pattern changes are visible. However unfortunately, no information on the patient's condition was recorded on the ward during these days.

\subsubsection{COVID-19 Patient with Complications}

In Fig.~\ref{fig:patient-complications}, the multimorbid 71-year-old man SARS-CoV-2 positive patient is shown, who developed complications during his infection. The patient was admitted to the COVID-19 ward on day 6. In addition to COVID-19, the patient had various neurological issues, which resulted in aspiration (stomach contents breathed into the lungs) on about day 10, after which he developed pneumonia.
In the visualization, this corresponds to the initially low HR and HRV (expected for COVID-19~\cite{COESB:20,NSH:20}). After a slow improvement with an increased in HR, his HR rose sharply due to the aspiration, his HRV fell again and correspondingly his RR increased significantly. After several days, the values normalized again as a sign that the pneumonia has subsided under antibiotic therapy, which started on day 11. Note: The determination of the point in time when the pneumonia can be considered to have subsided is ambiguous.  It could be defined as once the symptoms (fever, shortness of breath) had subsided, when the patient became active again or when there were no pathological findings visibile in X-rays approximately four weeks after the symptoms have receded. Nevertheless, prominent pattern changes are clearly visible a few days after the antibiotic therapy started.

\subsubsection{Stressed COVID-19 Patient} 

In Fig.~\ref{fig:patient-stressed}, the patient history of a 65-year-old man who experienced a stressful  situation is illustrated. The patient entered the hospital in a significantly reduced general condition (very weak and tired). He was first admitted to the emergency unit before he was transferred to the COVID-19 ward on day 3 as a result of a positive SARS-CoV-2 test. He received two liters per minute of oxygen through nasal application and intravenous fluid replacement and recovered within a few days. His HR increased over the course of his recovery, which might be related to the subsiding infection as well as to an increase in activity after about day 7. 
The prominent signal increases in HR and RR in the early afternoon of day 7 represent a particularly stressful  situation for the patient. This is shown as red in the heat map and as higher bars in the bar charts. The team on the medical ward documented a difficult discussion with the patient on that day.

\section{Summary}

In this study, we evaluated two effective visualizations for the primary use case of communicating relevant health patterns in wearable data between clinical staff and data scientists through visualizations. In our survey, the customized heat maps were  considered particularly useful for detecting extreme values and for identifying trends (challenge Sec.~\ref{sec:c_spot_patterns}); the bi-directional bar charts revealed themselves to be well suited for identifying repeating, changing or irregular patterns as the patient's health condition evolved (challenge Sec.~\ref{sec:c_spot_patterns}). The survey conducted on the heat map visualizations helped to improve the plotting layout and initiated the usage of bar charts. The final layout effectively reduced the amount of  data displayed by temporally aggregating data for selected vital signals (challenge Sec.~\ref{sec:c_data_ranges}), which were quality filtered (challenges Sec.~\ref{sec:c_missing}--\ref{sec:c_implausible}). The final visualizations clearly depict relevant information with annotations such as the midnight lines or the time axis labels to quickly identify periodic patterns and recognize time intervals with missing data (challenges Sec.~\ref{sec:c_color_coding}--\ref{sec:c_spot_patterns}). 
A lesson learned was that the carefully designed heat maps were viewed as overly sophisticated,  implying a need to train the care team if applied to other clinical use cases. Nevertheless, the developed and evaluated visualization layout can easily be transferred to other visualization types.

In the future, it would be interesting to visually explore hypotheses on circadian rhythms, which are clearly visible in the developed bar charts or to compare our results with SpiralGraphs~\cite{AMMST:08}. Finally, the next steps could be to integrate interactive elements into the visualizations and to deploy our visualizations to real-time patient monitoring.

The proposed wearable health data pipeline uses elements of visual data history illustrated in the dataset of 84 COVID-19 patients, who were monitored using wearables during the first pandemic wave in early 2020. The developed approach facilitated  communication between clinical staff and data scientists for the purpose of discovering noteworthy patterns in the evolution of the patients' health conditions. With the help of this exploratory visual process, we were able to efficiently select three exemplary COVID-19 patient histories.

Our approach was demonstrated for our primary use case and applied to wearables for which we described the specific data challenges in visualization. However, the approach can be extended to further clinical use cases such as the detection of critical patient situations or ward monitoring.

\section{Acknowledgments}
The icons in Fig.~\ref{fig:concept} were made by Freepik~\cite{Freepik:20}. The authors thank the participants of the conducted visualization survey, which resulted in significantly improved visualizations for the given use case. Thanks to Norman Juchler and Darren Mace for his valuable review on the manuscript.

\bibliographystyle{plain} 
\bibliography{5_bibliography}

\section{Appendix: Visualization Survey Details}
\label{sec:appendix}

\subsection{Questionnaire}

The following questions were answered in written by three persons from hospital (two medical doctors and one clinical project manager) and seven health data science researchers:
\subsubsection*{Q1: Which color scheme do you find most intuitive for the vital signs?}
\begin{enumerate}[label=\alph*.]
        \item Red-white-blue (RWB)~\cite{BHSWH:20}
        \item Red-yellow-blue (RYB)~\cite{BHSWH:20}
        \item Other from Colorbrewer 2.0~\cite{BHSWH:20}
\end{enumerate}

\subsubsection*{Q2: Do you prefer continuous or discrete color schemes for the visualization of the vital signs?}
\begin{enumerate}[label=\alph*.]
        \item Continuous
        \item Discrete
\end{enumerate}

\subsubsection*{Q3: How many discrete color steps do you prefer?}
\begin{enumerate}[label=\alph*.]
        \item 5 
        \item 7
        \item 9
\end{enumerate}

\subsubsection*{Q4: How useful are the visualizations?}
This question was adapted to our visualizations from the system usability scale (SUS)~\cite{Brooke:96,SUS:20}. In contrast to the original SUS, to save time for the reviewers, we skipped two questions and let reviewers only answer yes or no instead of using a Likert scale.
\begin{enumerate}[label=\alph*.]
        \item I think that I would like to use these visualizations frequently
        \item I find the visualizations unnecessarily complex
        \item I think that I would need the support of a technical person to use this visualization
        \item I thought there was too much inconsistency in this visualization
        \item I would imagine that most people would learn to use this visualization very quickly
        \item I found the visualization very cumbersome to use
        \item I feel very confident using this visualization
        \item I needed to learn a lot of things before I could get going with this visualization
\end{enumerate}

Skipped questions:
\begin{itemize}
    \item I thought the visualization was easy to use
    \item I found the various functions in this visualization were well integrated
\end{itemize}

\subsubsection*{Q5: What labeling of the time axis would be helpful?}
Free text answer

\subsubsection*{Q6: What do you particularly like about the heat maps?}
Free text answer

\subsubsection*{Q7: Where do you see problems with the heat map visualizations?}
Free text answer

\subsubsection*{Q8: What would you improve in the visualizations?}
Free text answer

Questions only posed to clinical staff:

\subsubsection*{Q9: How would you choose the value ranges for the vital signs?}
\begin{enumerate}[label=\alph*.]
        \item Individual absolute min/max values per signal for each patient
        \item The same absolute min/max values per signal for all patients
        \item If b. what specific range per signal?
\end{enumerate}

\subsubsection*{Q10: Are the visualizations clinically usable/helpful/relevant?}
Free text answer

\subsection{Answers}
The answers for questions Q1--Q4 are summarized in Tab.~\ref{tab:table1}.

\begin{table}[ht]
  \begin{center}
    \caption{Questionnaire answers by clinical persons (CP) and health data scientists (DS).}
    \label{tab:table1}
    \begin{tabular}{l|l|l|l|l} 
      \textbf{} & \textbf{Q1} & \textbf{Q2} & \textbf{Q3} & \textbf{Q4}\\
      background & colors & scale bar & steps & SUS\\
      \hline
      CP & RYG & discrete & 3 & b,f\\
      CP & RWB & discrete & 5 & a,c,e\\
      CP & RWB/RGB & continuous & 7+ & -\\
      DS & RWB & continuous & - & -\\
      DS & RWB & continuous & 7 & a,e,g\\
      DS & RWB & continuous & 7 & a,e,g\\
      DS & RYB & continuous & 9 & a,e\\
      DS & RYB & discrete & 5 & e,g\\
      DS & RYB & discrete & 7 & e\\
      DS & RYB & discrete & 7 & e\\ 
    \end{tabular}
  \end{center}
\end{table}

\subsubsection*{A5: time axis labeling}
Generally, the hourly labels were considered easily understandable. Noteworthy comments:
\begin{itemize}
    \item Write hours as 12:00 instead of 12
    \item Add day information in major ticks (two answers)
    \item Hours past from admission time point
    \item More time points, e.g., 00:00, 08:00 12:00, 18:00 (six answers)
    \item Start visualization at an intuitive time point such as 00:00
\end{itemize}

\subsubsection*{A6: what was liked about visualizations}
Noteworthy comments:
\begin{itemize}
    \item Intuitive (four answers)
    \item Good overview of patient condition over the course of days (two answers)
    \item Once the heat map visualizations are understood, the vital signs are easily visible at a glance
    \item Trends and extreme values clearly visible
    \item Useful to synchronize with medical events, e.g., medication
    \item Simple representation
    \item The information density
\end{itemize}

\subsubsection*{A7: seen problems/challenges}
\begin{itemize}
    \item Users must be trained -- otherwise the visualizations are not helpful
    \item Understanding color scale bars, since they are individual per signal (two answers)
    \item The color scale bars must be well matched to their medical relevance
    \item Loss of information for discrete color steps
    \item With too many colors the color-coded meaning is not clear
    \item Difficult to read exact specific values from color bars (two answers)
    \item For color-blind people it might not be obvious/intuitive to read color-codings -- even if the color shades are distinguishable
    \item Red color-coding of heart rate variability looks quickly very dangerous -- is this really clinically so dangerous?
\end{itemize}

\subsubsection*{A8: suggestions}
\begin{itemize}
    \item Make clearly visible where no data was recorded (mid-range white values vs. NAN grey values are difficult to distinguish)
    \item Labeling of time axis labels: add days and more time points (see A5)
    \item Make color scale bars larger
    \item Use spacing between signals
    \item Color coding: green=good, yellow=raised, red=critical
    \item The most important thing is to understand at a glance whether everything is ok. It would be good to have three colors only for everything is ok, borderline and too high/too low
    \item Interactive visualizations to display exact numeric values of specific values that a user hovers over
    \item Sometimes it is not the temporal value of a parameter that is relevant, but the cumulative/integrative effect of the value progression. For example, short peaks may be less interesting than longer-term deviations from the nominal value.
\end{itemize}

\subsubsection*{A9: value ranges}
For values such as heart rate, heart rate variability and respiration rate, the value ranges are heavily dependent on the individual patient. Additionally, there is a clear change in values during daytime and nighttime. For core body temperature and oxygen saturation, absolute values could be used. Typically, medical devices allow the expected value ranges to be configured individually for each patient.

\subsubsection*{A10: clinical relevance}

\begin{itemize}
    \item Yes, visualizations enable a faster (''at a glance``) assessment of the situation, which then leads to a patient being examined in more detail
    \item Yes, because patterns can be easily detected over time; helpful because major signals are visible at a glance; well aggregated time points; however, they are intellectually demanding
    \item Yes, they work clinically if generated with three colors only
\end{itemize}

\end{document}